\documentclass[superscriptaddress,floatfix,twocolumn,showpacs,preprintnumbers,amsmath,amssymb,aps,prl,longbibliography]{revtex4-1}
\usepackage{color}
\usepackage{subfigure}
\usepackage{graphicx}
\usepackage{bm}

\begin{document}

\title{Orbital angular momentum from semiconductor high-order harmonics}

\author{David Gauthier}
\author{Shatha Kaassamani}
\author{Dominik Franz}
\author{Rana Nicolas}
\affiliation{LIDYL, CEA, CNRS, Universit\'e Paris-Saclay, CEA-Saclay, 91191 Gif-sur-Yvette, France}
\author{Jean-Thomas Gomes}
\author{Laure Lavoute}
\author{Dmitry Gaponov}
\affiliation{Novae, ZI du Moulin Cheyroux, 87700 Aixe-sur-Vienne, France}
\author{S\'ebastien F\'evrier}
\affiliation{Universit\'e de Limoges, CNRS, XLIM, UMR 7252, 87000 Limoges, France}
\author{Ga\"etan Jargot}
\affiliation{Laboratoire Charles Fabry, Institut d’Optique Graduate School, CNRS, Universit\'e Paris-Saclay, 91127 Palaiseau, France}
\affiliation{Fastlite, 165 rue des Cistes P\^ole entreprise 95 06600 Antibes, Sophia Antipolis, France}
\author{Marc Hanna}
\affiliation{Laboratoire Charles Fabry, Institut d’Optique Graduate School, CNRS, Universit\'e Paris-Saclay, 91127 Palaiseau, France}
\author{Willem Boutu}
\author{Hamed Merdji}
\affiliation{LIDYL, CEA, CNRS, Universit\'e Paris-Saclay, CEA-Saclay, 91191 Gif-sur-Yvette, France}

\date{\today}

\begin{abstract}
Light beams carrying orbital angular momentum (OAM) have led to stunning applications in various fields from quantum information to microscopy. In this letter, we examine OAM from the recently discovered high-harmonic generation (HHG) in semiconductor crystals. HHG from solids could be a valuable approach for integrated high-flux short-wavelength coherent light sources. The solid state nature of the generation medium allows the possibility to tailor directly the radiation at the source of the emission and offers a substantial degree of freedom for spatial beam shaping. First, we verify the fundamental principle of the transfer and conservation of the OAM from the generation laser to the harmonics. Second, we create OAM beams by etching a spiral zone structure directly at the surface of a zinc oxide crystal. Such diffractive optics act on the generated harmonics and produces focused optical vortices with nanometer scale sizes that may have potential applications in nanoscale optical trapping and quantum manipulation.
\end{abstract}

\maketitle

Interaction between matter and photons carrying an orbital angular momentum (OAM) is nowadays an intensively studied topic in a large variety of applications, from quantum cryptography to the manipulation of particles \cite{Torres}. Light beams carrying OAM have a helical wavefront accompanied by a phase singularity in their center \cite{Padgett}. They show an azimuthal phase dependence $\exp(-i \l \phi)$, where $\l$ is the topological charge and $\phi$ is the azimuthal coordinate in the plane perpendicular to the beam propagation. Such beams carry an OAM of $\l\hbar$ per photon \cite{Allen}. The most common light beams carrying OAM display Laguerre-Gaussian modes characterized by a donut-shaped intensity profile. There is currently a large theoretical and experimental effort towards the generation of vortex beams with new coherent light sources \cite{Hernandez,Gauthier,Rebernik,Vieira,Petrillo}. Among the challenges involving OAM beams, increasing attention is devoted to the fundamental studies in highly nonlinear light-matter interactions. High-harmonic generation (HHG) based on frequency up-conversion of a high intensity Vis-IR femtosecond laser can generate coherent beams down to few nanometer wavelength and attosecond pulse duration \cite{Brabec}. This coherent light source is now very common and the underlying mechanisms are well understood for various generation media, particularly for atomic gases and plasma targets. Recently, in both media, the OAM of the emitted harmonics has been investigated. These studies confirmed the transfer and conservation of the OAM from the driving laser to the high harmonics \cite{Gariepy,Geneaux,Denoeud}.

Reported for the first time in 2011 \cite{GhimireNat}, HHG in solids, namely bulk semiconductor crystals, has triggered numerous studies focused on the understanding of the phenomenon, proposing different models and pushing experimental setups \cite{GhimireJPB,VampaJPB}. The harmonic generation mechanism is different from the dilute matter due to the high density, band structure and periodicity of the crystal. Moreover, it requires a relatively lower intensity compared to HHG in gases which makes it well suited for high repetition rates (multi MHz) nanoscale integrated light sources \cite{KFLee,Luu}. An all solid-state HHG source would open vast applications in science but also in industry. For example, it would be complementary to large scale facilities such as free-electron lasers, for experiments that require a low number of counts per pulse at high repetition rate to increase statistics. An interesting feature specific to the generation of high harmonics in solids is the possibility to tailor the generation medium; the solid can be patterned to control the process of generation and the emission of radiation \cite{Kim,Sivis}. In comparison to HHG in gases, this offers a significant additional degree of freedom. Such arrangement of the solid prior to the generation can be used to enhance the efficiency of the HHG at a nanometer scale \cite{Han,Sivis}. Moreover, the pulse properties can be shaped directly at the source of the HHG emission by changing the characteristics of the generation medium.

\begin{figure}[!h]
\includegraphics[width=85mm]{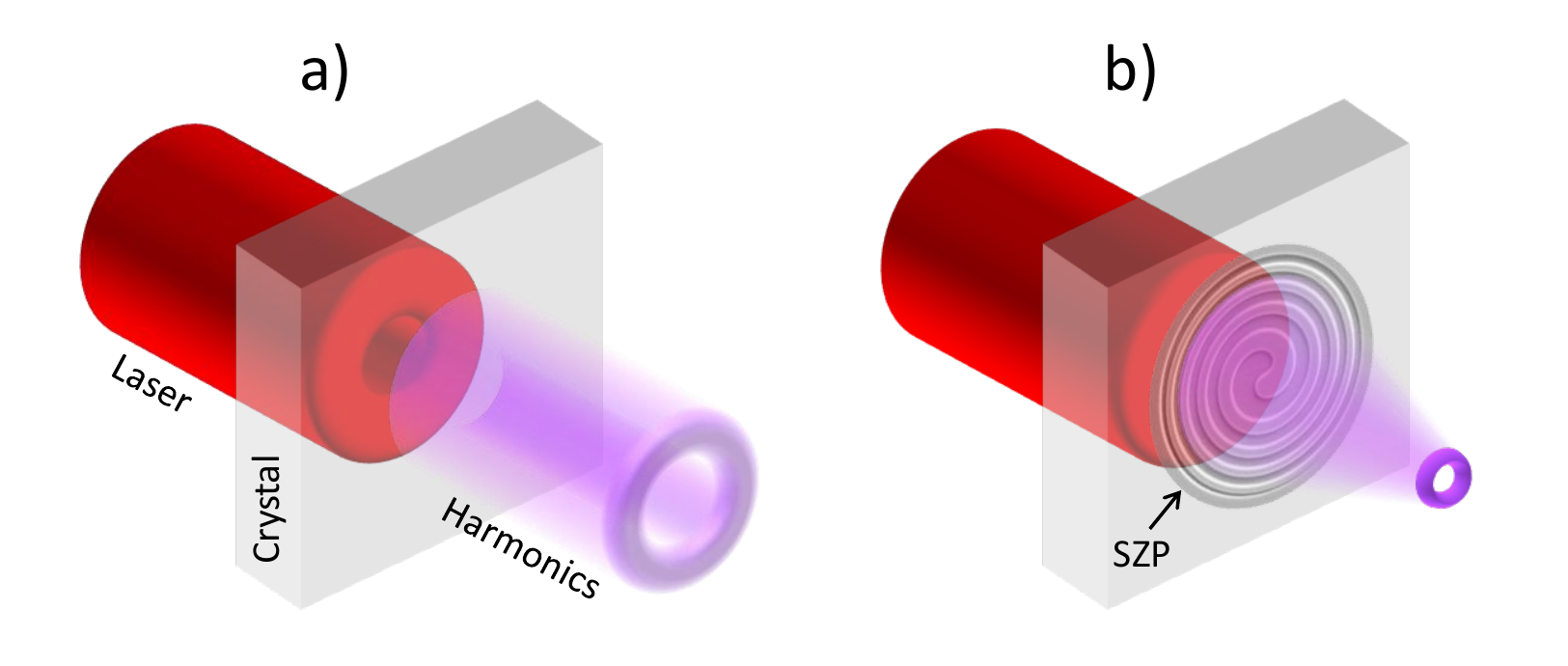}
\caption{Sketch of two approaches to shape the spatial profile of the generated harmonics in solids applied to the creation of a beam carrying an OAM: (a) the driving laser carries an OAM; (b) a spiral zone plate (SZP) is patterned at the surface of the generation crystal, acting as a diffractive optics on the harmonics.
\label{fig1}}
\end{figure}

In this letter, we investigate HHG beams in solids carrying OAM using two approaches as illustrated in figure~\ref{fig1}. The first case considers the generation by means of a Laguerre-Gaussian laser beam in a bare crystal (Fig.~\ref{fig1} (a)). Similarly to HHG in gases, we show that the OAM of the harmonic beams is conserved such that their topological charge $\l_{q}$ is a multiple of the harmonic order $q$ following the relation $\l_{q} = \l_{l} \times q$, with $\l_{l}$ the topological charge of the driving laser. The second case exploits the solid state nature of the generation medium to macroscopically control the harmonic beams. By patterning the surface of the crystal one can design either refractive or diffractive optics. For example, nanostructured surfaces can be designed and arranged spatially in such a way that the harmonics generated in the bulk diffract and build a new mode of the light. Figure~\ref{fig1} (b) shows a diffractive optic, specifically a spiral zone plate (SZP) \cite{Heckenberg,Bazhenov,Sakdinawat}, etched at the surface of the crystal. The SZP combines beam focusing, just like an ordinary Fresnel zone plate, together with the OAM mode conversion to produce focused optical vortices carrying an arbitrary topological charge.

\begin{figure}[!h]
\includegraphics[width=85mm]{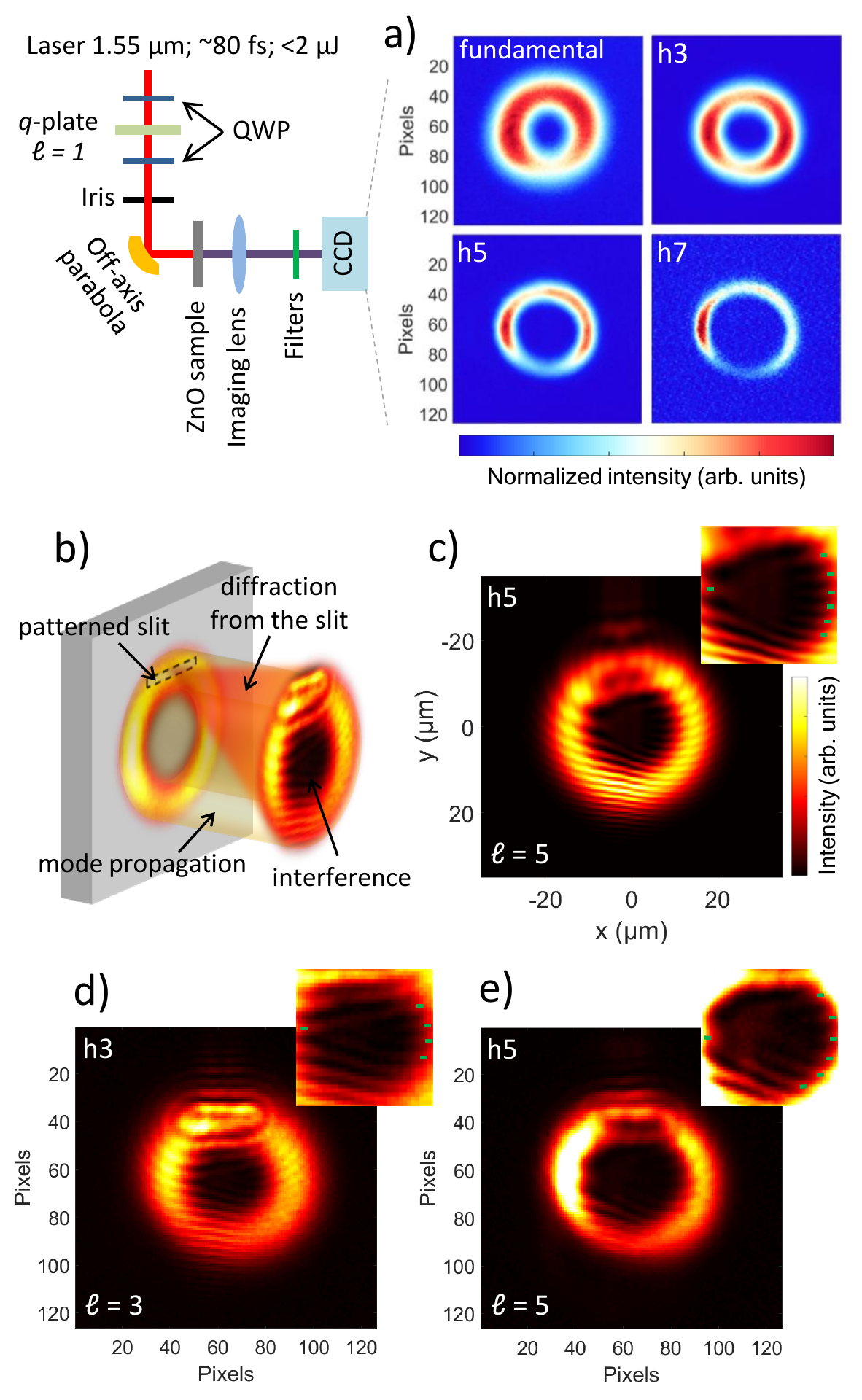}
\caption{Generation with a laser carrying an OAM. Sketch of the experimental setup; a \textit{q}-plate between two quarter waveplates (QWP) imparts an OAM with a topological charge $\l = 1$ on the laser. The beam is then focused by a parabolic mirror in a ZnO crystal. The generated harmonics are collected by a lens to image the mode at the exit of the crystal (a). Each harmonic order is selected using bandpass filters. (b) The topological charge of the modes is defined using self-referenced interferences. (c) Simulation with the experimental parameters of the self-referenced interferogram for harmonic h5. (d) and (e), experimental results for harmonics h3 and h5. The insets show the detail of the fork pattern. The green dots highlight the extremity of each fringe.
\label{fig2}}
\end{figure}

The first experiment was performed with an optical parametric chirped pulse amplification laser system emitting at 1550 nm pumped with a high-energy femtosecond ytterbium-doped fiber amplifier \cite{Rigaud}. It provides pulses with duration of $\approx 80\:\text{fs}$ and energy of $\approx 2\:\mu\text{J}$ at the repetition rate of $125\:\text{kHz}$. The laser beam is converted from a Gaussian to a Laguerre-Gaussian mode, with a unitary topological charge, using a \textit{q}-plate between two quarter waveplates \cite{Marrucci}. The laser is focused by a $5\:\text{cm}$ focal length off-axis parabola ($\text{f}/15$) in a $500\:\mu\text{m}$ thick zinc oxide (ZnO) crystal (laser propagation along the optical axis of the crystal) reaching intensities ranging from $0.2$ to $1.5\:\text{TW/cm}^{2}$. An aspheric lens (numerical aperture $\text{NA} = 0.63$) and a CCD camera are used to image the harmonic beams at the exit of the crystal (see figure~\ref{fig2}). Figure~\ref{fig2} (a) displays the modes at the exit of the crystal for the fundamental and the generated harmonics h3 (517 nm), h5 (310 nm) and h7 (221 nm). The modes show the expected ring-shaped distribution. The radius of the modes is estimated to be about $15\:\mu\text{m}$. The thickness of the rings decreases with an increase of the harmonic order, as a consequence of the nonlinear response to the annular fundamental intensity distribution in the sample. The imperfections in the fundamental mode are enhanced for high harmonics, and suggest a decrease of the Laguerre-Gaussian mode purity. To determine the topological charge of each mode we performed self-referenced spatial interferometric measurements \cite{DGLee}. The principle is illustrated in figure~\ref{fig2} (b). A rectangular slit ($15 \times 1\:\mu\text{m}$ size) patterned on the crystal front surface diffracts a part of the harmonic beam generated in the crystal. At a distance of about $100\:\mu\text{m}$ from the crystal, the diffraction spreads over the harmonic beam (not affected by the slit) that propagates outside the crystal. The slit is small enough with respect to the OAM mode such that the diffracted wavefront does not contain any phase singularity and can be approximated by a spherical wave front. Interference of the latter with the helical wave front of the main beam gives a characteristic fork grating pattern, as depicted in the simulation (figure~\ref{fig2} (c)). The topological charge of the mode is deduced directly from the number of bifurcations. Figures~\ref{fig2} (d) and (e) display the experimental results for the 3rd and 5th harmonics, respectively. In figure~\ref{fig2} (d), one fringe splits in 4 fringes which gives a topological charge of 3. For the 5th harmonic, one fringe splits in 6 corresponding to a topological charge of 5. These results validate the \emph{multiplicative} rule for the OAM transfer from the generation laser to the harmonics. They have been obtained with laser peak intensity in vacuum estimated to $0.8\:\text{TW/cm}^{2}$ that corresponds to a Keldysh parameter near the tunneling limit of strong-field ionization \cite{GhimireJPB}. We conclude that in this highly nonlinear regime, the mechanisms related to HHG in semiconductors do not affect the conservation rules of OAMs for harmonics having photon energies below as well as above the ZnO bandgap (3.3 eV).

\begin{figure}[!h]
\includegraphics[width=85mm]{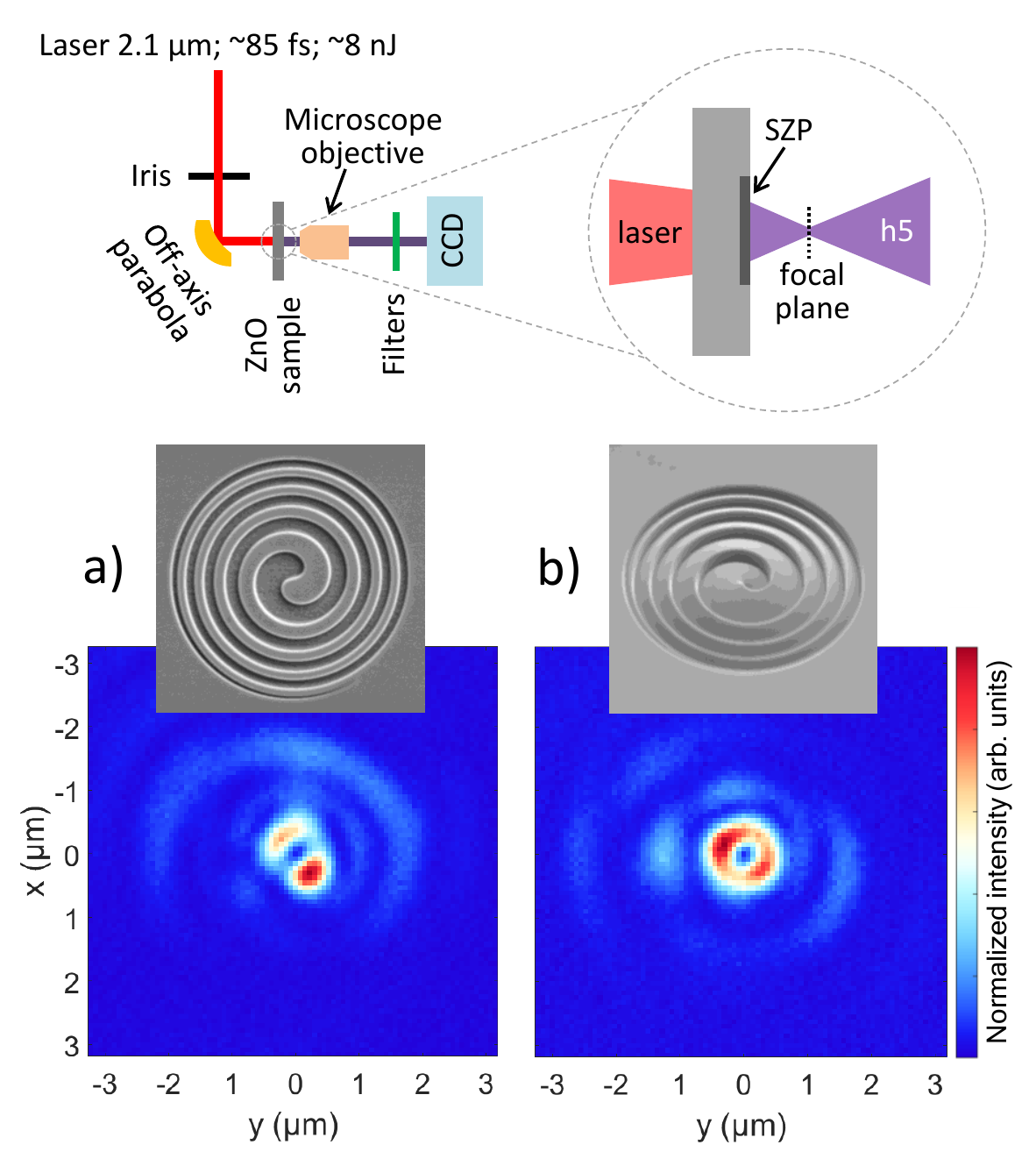}
\caption{Spatial beam shaping using a SZP that combines beam focusing and OAM manipulation. Sketch of the experimental setup; a microscope objective is used to image the focal plane of the SZP. Experimental results for a binary (a) and a non-binary (b) SZP etched at the surface of a ZnO crystal. Scanning electron microscope images of each diffractive structure (diameter $=10\:\mu\text{m}$, normal incidence (a) and perspective view (b)) and resulting intensity distributions at the focal plane for harmonic h5.
\label{fig3}}
\end{figure}

\begin{figure}[!h]
\includegraphics[width=85mm]{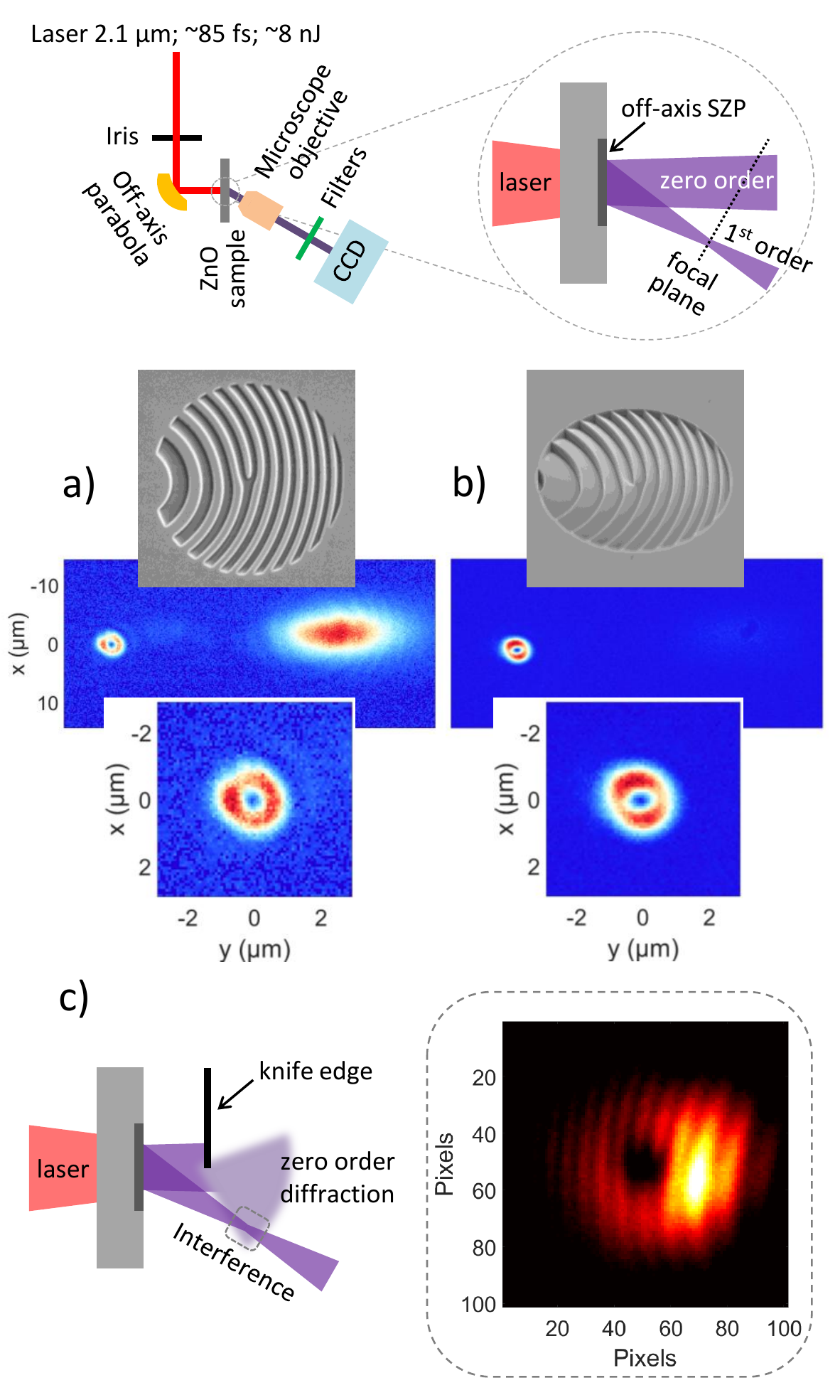}
\caption{Off-axis SZP. Sketch of the experimental setup; the microscope objective is tilted by an angle of 31 degrees to be aligned along the optical axis of the first diffraction order. Experimental results for a binary (a) and a non-binary (b) off-axis SZP. Scanning electron microscope images of each diffractive structure (diameter $=10\:\mu\text{m}$) with the resulting intensity distributions at the focal plane of the first order, showing the zeroth and first orders of diffraction. The first order is magnified in the inset. The zeroth order shows strong aberrations due to the coupling with the microscope objective. (c) Sketch of the procedure used to measure the topological charge by interference of the first order with the zeroth order and the resulting fork grating.
\label{fig4}}
\end{figure}

The second experiment was performed with a commercial all-fiber (oscillator and amplifier) femtosecond mode-locked laser emitting at 2100 nm \cite{Novae}. It provides pulses with duration of $\approx 85\:\text{fs}$ and energy of $\approx 8\:\text{nJ}$ at a repetition rate of $18.66\:\text{MHz}$. The laser beam is directly focused at the rear side of the sample by a $2.5\:\text{cm}$ focal length off-axis parabola ($\text{f}/6$) at a peak intensity of $0.3\:\text{TW/cm}^{2}$. The sample is a ZnO crystal with various spiral zone plates etched at its surface using a nano-focused ion beam. The image of the beam diffracted by the SZP is obtained with an objective lens (numerical aperture $\text{NA} = 0.65$) coupled to a CCD camera (see figure~\ref{fig3}). The desired SZPs are pure phase objects computed numerically by combining the phase profiles of a helical and spherical waves following the relation $P = \exp(-i \l \phi) \exp(i 2\pi \sqrt{R^{2}+r^{2}}/\lambda)$, where $\lambda$ is the wavelength of the spherical wave of radius of curvature $R$ (determining the focal length of the SZP), and $r$ is the radial coordinate in the plane perpendicular to the beam propagation. All the SZPs have been designed to generate optical vortices with a topological charge $\l = 1$. The calculations are made without paraxial approximation to take into account the high numerical aperture of the SZPs. From the computed phase $P$, two types of etching profile have been defined: the \emph{non-binary} SZP directly corresponds to the wrapping of $P$ within $0-2\pi$, and the \emph{binary} SZP is the binarization of the latter at $0$ and $\pi$. For below bandgap energies, the harmonic radiation is generated in the bulk and the etching depth $d$ in the crystal can be defined as $d = \lambda P/2\pi(n-1)$, where $n$ is the index of refraction of the ZnO material at the relevant wavelength. For the data in Fig.~\ref{fig3} and \ref{fig4} showing harmonic h5 (420 nm), $n\approx 2.1$.

Figure~\ref{fig3} (a) shows the binary SZP and the intensity distribution measured at the focal plane of the first diffraction order located at $6\:\mu\text{m}$ from the sample. Although imperfect, the mode shows the characteristic donut-shaped profile. The main cause of the imperfection is the restricted number of illuminated zones (grooves) of the SZP that limits an efficient build-up of the diffracted wave. Indeed, only the central part of the structure is actually illuminated by the $3\:\mu\text{m}$ diameter (FWHM) harmonic beam generated in the bulk. Increasing the number of illuminated zones would imply to increase the size of the laser beam or decrease the focal length of the SZP. However, the design of the SZP was a tradeoff between the laser energy available and the image resolution allowed with our objective lens. Another source of imperfection is coming from the fabrication defaults; the etched depth is about 205 nm, inducing a phase step about $7\%$ larger than requested. Moreover, the gallium ions contamination and the ruggedness induced by the etching process with the focused ion beam might have modified the transmittance of the optics. The last source of imperfection is coming from residual aberrations in the generating laser and misalignments. Figure~\ref{fig3} (b) shows the design of the non-binary version of the previous SZP and the resulting mode. This non-binary SZP is rigorously not a diffractive optics because it is based on the principle of a Fresnel lens. For this reason, it does not suffer from the problem of the limited number of illuminated zones, and the donut-shaped profile, with a radius of 380 nm, is better defined. However, due to small errors in the etching depth, it also shows a diffractive behavior with presence of diffraction orders, which explains the parasitic signal surrounding the main mode similar to the binary SZP.

Figure~\ref{fig4} displays results obtained with a binary and a non-binary off-axis SZP. The computed pattern of off-axis SZPs is achieved by a transversal shift between the centers of the spherical and helical waves. The shift is of $6\:\mu\text{m}$ for a radius of curvature of $10\:\mu\text{m}$. Consequently, the first order of diffraction is tilted from the main optical axis by an angle of 31 degrees (see sketch figure~\ref{fig4}). Fig.~\ref{fig4} (a) shows the etched binary off-axis SZP and the intensity distribution in the focal plane of the first order of diffraction. Compared to the previous binary on-axis SZP, the mode quality is significantly improved because the off-axis design leads to a larger density of zones illuminated in the center of the optics. Off-axis SZPs have other advantages: the diffraction orders are spatially separated, therefore there is no parasitic signal surrounding the main mode; for properly chosen parameters, beams from different harmonic orders can be split as well. The latter is particularly interesting for fully integrated and compact sources of optical vortices. For the non-binary off-axis SZP (figure~\ref{fig4} (b)), the blazed grating profile optimizes the flux diffracted in the first order, which contains $76\%$ of the detected photons, when this number is only of $9\%$ for the binary case. Finally, we performed interferometric measurements in order to verify the topological charge of the beam shaped by the SZP. In this case, the reference beam is provided by the zero order that spreads over the first order by means of a knife-edge, as depicted in figure~\ref{fig4} (c). We find a fork grating pattern with one bifurcation, which confirms the unitary topological charge of the OAM beam.

In conclusion, we reviewed and demonstrated two different methods to generate high-harmonic photons carrying an OAM from a semiconductor crystal. First, the conservation rule of OAM beam is verified for below and above bandgap harmonics in the strong-field regime of interaction. Then, we show that integration of a SZP to the generation medium produces focused optical vortices with nanometer scale sizes. For example, such beams can lead to strong localized forces and torques creating optical tweezers \cite{He}. Structured surfaces have been established for below bandgap harmonics. Shaping the crystal surface can change the intensity distribution of the driving laser with the appearance of local field confinements that modify the properties of the generated harmonics and possibly enhance the emission \cite{Sivis}. This is particularly relevant for above bandgap harmonics that are not efficiently generated in the bulk crystal due to absorption. Therefore, structured surfaces for above bandgap harmonic generation have to consider together the action on the generating laser and on the harmonics. The resulting 2-in-1 design can combine enhancement and spatial shaping of the harmonic emission. Pushing the idea of integrated devices can also merge with other nano-technologies such as nano-plasmonics \cite{Vampa}. For example, the generating medium could be arranged as a metamaterial. Planar devices based on metasurfaces enable various manipulations beyond the simple light diffraction, including the control of the polarization (spin) and orbital angular momentum of light \cite{Yu}. HHG in solids would take advantage of the numerous possibilities offered by the metasurfaces for the control of the light. 
\\
\\
We acknowledge financial support from the French ministry of research through the 2014 IPEX, 2016 BISCOT, 2017 PACHA, and LABEX PALM (ANR-10-LABX-0039-PALM) through the grants Plasmon-X and HILAC. We acknowledge the financial support from the French ASTRE program through the NanoLight grant and the support from the DGA RAPID program through the 2017 SWIM LASER grant.
\\
\\
\end{document}